\documentclass{emulateapj} 
\usepackage{graphicx}
\usepackage{color,calc}
\usepackage{amsmath}
\usepackage{amssymb}
\usepackage{amstext}

\begin{document}

\title{Evolution of the Binary Fraction in Dense Stellar Systems}
\shorttitle{Evolution of the Binary Fraction in Dense Stellar Systems}
\submitted{submitted to ApJ} 
\author{John M. Fregeau\altaffilmark{1,2,3}, Natalia Ivanova\altaffilmark{4}, and Frederic A. Rasio\altaffilmark{5}}
\shortauthors{FREGEAU, ET AL.}
\altaffiltext{1}{Kavli Institute for Theoretical Physics, UCSB, Santa Barbara, CA 93106}
\altaffiltext{2}{fregeau@kitp.ucsb.edu}
\altaffiltext{3}{Chandra/Einstein Fellow}
\altaffiltext{4}{Department of Physics, University of Alberta, Edmonton, AB, T6G 2G7, Canada}
\altaffiltext{5}{Department of Physics and Astronomy, Northwestern University, Evanston, IL 60208}

\begin{abstract}
Using our recently improved Monte Carlo evolution code, we study the evolution of
the binary fraction in globular clusters.  In agreement with previous $N$-body
simulations, we find generally that the hard binary fraction in the core tends to increase 
with time over a range of initial cluster central densities for initial binary
fractions $\lesssim 90\%$.  The dominant
processes driving the evolution of the core binary fraction are mass segregation of
binaries into the cluster core and preferential destruction of binaries there.
On a global scale, these effects and the preferential tidal stripping of single stars
tend to roughly balance, leading to overall cluster binary fractions that are roughly 
constant with time.  Our findings suggest that the current hard binary fraction 
near the half-mass radius is a good indicator of the hard primordial
binary fraction.  However, the relationship between the true binary fraction and the fraction
of main-sequence stars in binaries (which is typically what observers measure) is non-linear
and rather complicated.  We also consider the importance of soft binaries, which
not only modify the evolution of the binary fraction, but can drastically change
the evolution of the cluster as a whole.  Finally, we describe in some detail the recent addition
of single and binary stellar evolution to our cluster evolution code.
\end{abstract}


\section{The Binary Fraction}\label{sec:binfrac}
Observations and recent theory strongly suggest that the initial mass function (IMF)
is universal among non-zero metallicity stars \citep[e.g.,][]{2003PASP..115..763C}.
Indeed, \citet{2008arXiv0811.1035B} suggests that radiative feedback may naturally
regulate the star formation process so as to produce an IMF that is only weakly
dependent on the properties of the progenitor molecular cloud.  Naively, one
would also expect that other features of the initial stellar population---like
the binary fraction---should be nearly universal.  Hydrodynamical star
formation simulations yield companion star frequencies and binary
fractions that are largely independent of the properties of the progenitor
molecular cloud (although the statistics in some cases are marginal), 
and are quite consistent with observations 
\citep{2008arXiv0811.0163B,2003MNRAS.339..577B,2005MNRAS.356.1201B}.

Observations of stars in low stellar density environments
where dynamics is unimportant, such as the solar neighborhood, 
yield a binary fraction of $\sim 50\%$ among solar-type stars, 
with an increasing trend with primary mass
\citep[e.g.,][]{1991A&A...248..485D,1992ApJ...396..178F}.
Open clusters similarly show such large binary fractions
\citep{1996AJ....112..628F}.
However, observations of dense globular cluster cores typically yield
binary fractions that are significantly smaller.
{\em HST} observations of the core-collapse cluster NGC 6397
yield a binary fraction of $\approx 5\%$ in the core and
$\approx 1\%$ beyond the half-mass radius \citep{2008AJ....135.2155D}.
For the canonical non core-collapse cluster 47 Tuc,
the binary fraction is $\approx 13\%$ \citep{2001ApJ...559.1060A}.
The core binary fraction generally ranges from a few percent
to tens of percent, approaching $50\%$ in some cases for less dense clusters
\citep{2007MNRAS.380..781S}.
Where measured, the binary fraction outside the core is always smaller
\citep[see table in][]{2008AJ....135.2155D}.
The question naturally arises: Are the currently observed relatively
small core binary fractions in globular clusters consistent with 
initially larger binary fractions of $\sim 50\%$?

There are many strongly coupled processes that determine the evolution of the core binary fraction
in a dense stellar system.  Stellar evolutionary processes alone can affect the properties of a
binary greatly, causing it to expand or shrink via mass transfer or winds, circularize 
via dissipative effects, lose mass, receive a systemic velocity kick due to a supernova, or 
disrupt or merge.  The properties of the binary feed into the dynamical
interaction rate with other stars or binaries, causing it to interact more or less
frequently depending on its semimajor axis, eccentricity, mass, and systemic velocity.  
A strong dynamical interaction of a binary can disrupt it, exchange one of its members
for an incoming star, cause
its orbit to expand or shrink, modify its eccentricity, increase its systemic
velocity via gravitational recoil, or cause two or more stars to physically collide.
The dynamically modified binary properties feed back into binary stellar evolution,
possibly initiating or halting mass transfer, or increasing tidal effects.
In contrast to stellar evolutionary processes, the dynamical interaction rate 
depends on the cluster density and velocity dispersion, which evolve with time.
Since binaries are typically more massive than single stars, mass segregation can increase their numbers in the core
at the expense of single stars.  The tidal effects of the host galaxy will
preferentially strip single stars from the halo of the cluster.  

For a globular cluster of typical mass ($\sim 10^5 \, M_\sun$) and size (half-mass radius $r_h \sim 3 \, {\rm pc}$),
its global evolution can be divided into three phases according to the timescales of the relevant
physical processes.  At early times ($\sim {\rm few} \times 10$ Myr), the evolution is largely driven by stellar evolutionary
mass loss from the most massive stars in the cluster.  
At intermediate times ($\sim$ few Gyr), as mass loss from stellar evolution has slowed, the evolution
is driven primarily by two-body relaxation.  At late times \citep[possibly beyond a Hubble time;][]{2007MNRAS.379...93H},
when the core has reached sufficiently high density for binaries to strongly interact dynamically
and release enough energy to prevent core collapse, the properties of the cluster are determined
by the makeup of the binary population in this quasi-equilibrium ``binary burning'' phase.  

The core binary fraction is clearly a quantity that is affected by nearly all physical processes
operating in a cluster, and is of obvious observational interest.  Comparing observed
core binary fractions with simulation results (in combination with other observables)
is thus a good measure of our theoretical understanding of cluster evolution.
There can be dramatic differences in definition between 
the observed binary fraction and what theorists call the binary fraction, however.  

When measured
with the common offset main-sequence (MS) method, MS-MS binaries are detected by their appearance
as distinctly brighter MS objects.  The observed binary fraction is defined as the ratio
of the number of these ``binary sequence'' objects to the total number of objects in the MS
and the binary sequence, corrected for the assumed number of binaries with mass ratio 
so small they would blend in with the MS.  

The theorists' definition of the binary fraction
is typically the ratio of the number of binaries to the total number of ``objects'' (single
stars or binaries).  Furthermore, computational theorists tend to consider only ``hard''
binaries.  That is, binaries with binding energy greater than the typical particle energy, 
which typically become more tightly bound (harden) as a result of encounters
\citep{2003gmbp.book.....H}.  Soft binaries---binaries with binding energy less
than the typical particle energy in a cluster, which typically become less tightly
bound (soften) or dissociate completely---are less frequently considered.  We consider
in detail the difference between the observational and theoretical definitions of the 
binary fraction below, as well as the importance of soft binaries.

Recently, two very different simulation methods have been used to study the 
evolution of the binary fraction.  \citet{2005MNRAS.358..572I} have developed
a simplified Monte Carlo method in which a dense, massive cluster is modeled as a constant-density core
plus halo (to simulate the long-lived binary burning phase that clusters may reach late
in their evolution).  
Binaries and stars are evolved via the population synthesis
code StarTrack \citep{2008ApJS..174..223B}, and the strong dynamical interactions of binaries are
integrated numerically with Fewbody \citep{2004MNRAS.352....1F}.  Objects move between the core
and the halo due to mass segregation and systemic velocity changes resulting from dynamical encounters.
In this approach the core mass increases slowly with time, with very few stars leaving the core
after mass segregating into it.

\citet{2005MNRAS.358..572I} find, generally,
that the core binary fraction decreases significantly with time.  Even for a modest core density of 
$10^3\,{\rm pc}^{-3}$, they find that an initial binary fraction of 100\% yields
a core binary fraction of 27\% at 14 Gyr.  For the density of 47 Tuc, they find
that a 100\% initial binary fraction yields an 8\% core binary fraction at 14 Gyr.
It should be noted, however, that these figures include substantial numbers of {\em soft}
binaries---binaries that are so wide they are quickly destroyed by dynamical encounters.
If only the hard binaries in these simulations are counted, an initial binary fraction 
of 25\% in a $10^3\,{\rm pc}^{-3}$ core density cluster yields a 15\% core binary fraction
at 14 Gyr.  For a density of $10^5\,{\rm pc}^{-3}$ the core binary fraction evolves from
an initial 25\% to 7\%.

\citet{2007ApJ...665..707H} have used a direct $N$-body method, coupled with
the BSE single and binary stellar evolution routines 
\citep{2000MNRAS.315..543H,2002MNRAS.329..897H}, to study the evolution 
of the binary fraction.  The great benefit of this method is that it makes no simplifying
assumptions about the underlying cluster evolution.  On the other hand,
it is computationally expensive, currently limiting its application to open clusters
or globulars with low initial binary fractions.  \citet{2007ApJ...665..707H} 
find that the core (hard) binary fraction generally increases with time.  For a cluster of
$5 \times 10^4$ stars with a central density of $\sim 10^{3.5}\,{\rm pc}^{-3}$,
the core binary fraction rises from an initial 20\% to 52\% at 9 Gyr.  For lower
initial densities the degree of increase of the core binary fraction is similar.

\begin{figure}
  \begin{center}
    \includegraphics[width=0.9\columnwidth]{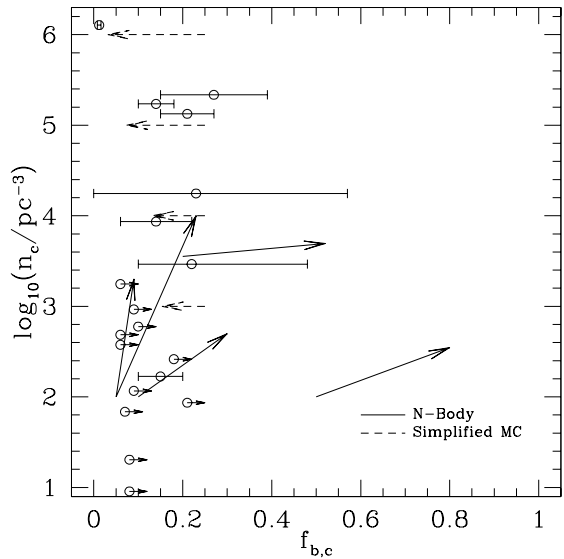}
    \caption{Evolution of $N$-body \citep[e.g.,][]{2007ApJ...665..707H} and 
      simplified Monte Carlo \citep{2005MNRAS.358..572I}
      cluster models in core number density---binary fraction space.
      Each model's evolution is represented as a simple arrow, with the tip at the final
      properties, and the tail at the initial properties.  For simplified
      Monte Carlo the final properties are measured at an age of 14 Gyr.
      For $N$-body they are measured at $\sim 15\,{\rm Gyr}$ in most cases,
      with the 50\% initial binary fraction model measured at 4 Gyr, 
      and the $10^{3.5}\,{\rm pc}^{-3}$ core density model measured
      at 9 Gyr.  Note that the binary fractions plotted here include only
      hard binaries.  For reference, we plot as open circles the current observed properties for 
      several Galactic globular clusters where measurement is possible, with data taken from the 
      table in \citet{2008AJ....135.2155D}.\label{fig:ivanovacomp}}
  \end{center}
\end{figure}

On the face of it, the discrepancy between the two methods appears irreconcilable.
However, the two methods operate at very different core densities and cluster masses, 
both of which affect the half-mass relaxation time and hence the mass segregation timescale,
as well as the binary dynamical interaction rate.  Fig.~\ref{fig:ivanovacomp} shows
the evolution of the various models in core number density---binary fraction space.
Note that the binary fractions plotted here include only hard binaries.  
Each model's evolution is represented as a simple arrow, with the tip at the final
properties, and the tail at the initial properties.  It is clear from this figure that
the two methods represent very different regions of parameter space, and could simply
be displaying different aspects of the same underlying physics.  The only point of concern
is the $N$-body model starting at $\sim 10^{3.5}\,{\rm pc}^{-3}$ and evolving toward a higher
binary fraction, nestled between two Monte Carlo models evolving in the opposite direction.

To elucidate the evolution of the binary fraction, and to address the discrepancy
between the existing $N$-body and simplified Monte Carlo models, we have performed
a grid of simulations with our newly-upgraded Monte Carlo cluster evolution code.
Note that our Monte Carlo code is very different from that of \citet{2005MNRAS.358..572I}.
While their code assumes a constant core density with time and samples binary interactions
using Monte Carlo techniques, our code self-consistently models the global evolution of a cluster,
using Monte Carlo techniques to sample the stellar distribution function when applying 
the effects of two-body relaxation.  The naming clash is the 
unfortunate consequence of the popularity and applicability of Monte Carlo techniques in general.

\section{Modern Simulations}\label{sec:modernsims}

\begin{figure}
  \begin{center}
    \includegraphics[width=0.9\columnwidth]{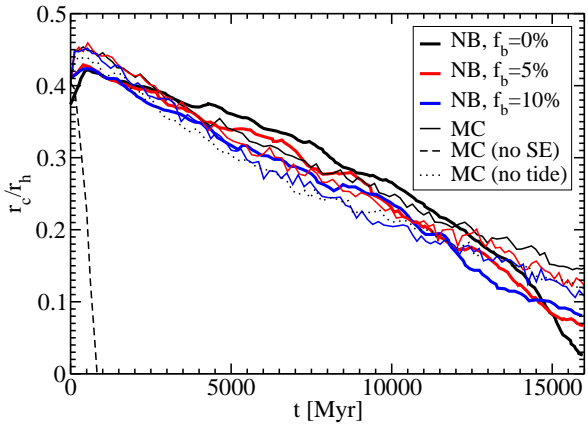}
    \caption{Evolution of the core to half-mass radius ratio for
      $N=10^5$ initial models with 0\%, 5\%, and 10\% primordial binaries, 
      comparing our new MC results with those of direct $N$-body 
      \citep{2007MNRAS.379...93H}.  The thick solid lines show the $N$-body models,
      with color denoting initial binary fraction, $f_b$.  The thin solid lines show our MC
      simulation with all relevant physics turned on (stellar evolution in singles
      and binaries, physical collisions, binary interactions, and a tidal boundary), again
      with color denoting the initial binary fraction.
      For the sake of comparison, MC simulations with stellar evolution turned off, and
      without a tidal boundary are shown in the thin dashed and dotted lines, respectively.
      For clarity, only the 0\% initial binary fraction runs for these comparison models are shown,
      since the 5\% and 10\% do not differ appreciably from the 0\% case.
      Clearly the evolution of this model is driven primarily by the effects
      of stellar evolution.  With the exception of the increased expansion of the
      cluster core at early times in the MC model, there is very good agreement between
      MC and $N$-body for all three binary fractions considered, suggesting that the 
      implementation of stellar evolution in the MC code is consistent with that
      of $N$-body.  Note that for the sake of comparison, the core radius here 
      is calculated using the standard definition for $N$-body simulations \citep{1985ApJ...298...80C}.\label{fig:hurleycomp}}
  \end{center}
\end{figure}

Our Monte Carlo (MC) code self-consistently models the evolution of star clusters
due to the effects of two-body relaxation, evaporation through a Galactic tidal
boundary, dynamical scattering interactions
of binaries, physical stellar collisions, and now single and binary stellar evolution.
The details of the method and its implementation are described in detail elsewhere
\citep{2000ApJ...540..969J,2001ApJ...550..691J,2003ApJ...593..772F,2007ApJ...658.1047F}.
Here we focus on the addition of stellar evolution.  

For coding simplicity and
for more directed comparisons with existing $N$-body simulations, we have
incorporated the BSE single and binary stellar evolution routines in
our Monte Carlo code \citep{2000MNRAS.315..543H,2002MNRAS.329..897H}.
In our code stellar evolution is performed for each object (single star
or binary) during a timestep in step with dynamics.  Since at early times
a cluster can lose a lot of mass due to supernovae, we make sure to limit
the timestep so that no more than a small fraction of the total cluster
mass is lost in one step (typically we set this fraction to $10^{-3}$).

To test that our inclusion of the stellar evolution routines
is accurate, we have compared with the $N$-body results of 
\citet{2007MNRAS.379...93H}, who evolved $N=10^5$ cluster
models with binary fractions ranging from 0 to 10\%.  The results
are shown in Fig.~\ref{fig:hurleycomp}, which displays the evolution
of the core to half-mass radius ratio ($r_c/r_h$) with time.  The data
from \citet{2007MNRAS.379...93H} were extracted from that paper using
ADS's Dexter applet \citep{2001ASPC..238..321D}.  For reference,
we also plot the evolution of a model with stellar evolution turned 
off, and a model without an external tidal field.  Since the model without
stellar evolution reaches core collapse in under 1 Gyr, and
since the model with no tide differs only minimally from the models
with all physics turned on, it's clear that stellar evolution drives
the evolution of the cluster.  These models are thus a good test of our
treatment of stellar evolution.  The agreement with $N$-body is quite
good given the vastly different methods, although the Monte Carlo models 
tend to expand more at early times due to supernovae.  The peculiar feature
that the evolution doesn't appear to depend strongly on initial binary fraction
is reproduced in our models.  At late times ($\gtrsim 15\,{\rm Gyr}$) our
models begin to diverge with $N$-body.  This is likely due to the fact
that the clusters have lost roughly 70\% of their stars by this time, and
our apocenter-based treatment of the tide will tend to underestimate
tidal mass loss as the number of cluster stars decreases, when
an energy-based criterion is more appropriate \citep[e.g.,][]{2008MNRAS.388..429G}.

In a forthcoming paper we will perform more detailed comparisons with existing
models of the open cluster M 67 and the globular clusters M 4 and NGC 6397 in the literature
\citep{2005MNRAS.363..293H,2008MNRAS.388..429G,2008MNRAS.389.1858H,2009arXiv0901.1085G}.
Given the vast differences between the $N$-body method and our Monte Carlo
method, we take the agreement between our models and those of \citet{2007MNRAS.379...93H}
in Fig.~\ref{fig:hurleycomp} as a sign that our implementation of BSE
in our code is at least consistent with that in $N$-body.

There is one aspect of our method that deserves special mention, however.
It is generally believed that if a cluster avoids a collisional 
runaway phase \citep[e.g.,][]{2006MNRAS.368..121F,2006MNRAS.368..141F} the
stellar-mass black holes formed early in a cluster's lifetime will quickly
sink to the core and dynamically decouple from the rest of the cluster,
undergoing their own evolution, much like an independent small star
cluster \citep{1993Natur.364..423S}.  The BH subsystem will quickly dissolve
through its own internal dynamics, ejecting all but one or two of the BHs
on a timescale $\lesssim 1\,{\rm Gyr}$.  Aside from removing nearly all
BHs from the cluster, the result is a mild energy injection into the
cluster core, causing it to expand somewhat at early times \citep{2007MNRAS.379L..40M}.
A typical star cluster of $N=10^6$ objects will contain a subsystem of up to
$\approx 10^{-3}N = 10^3$ BHs evolving independently in the core \citep{2006ApJ...637..937O}.
For the $N=10^5$ clusters considered in this work, the number is $\approx 100$.
The Monte Carlo method is not designed to handle subsystems of less than a 
few hundred objects, since they are often far from spherically symmetric, and large angle
scattering dominates.  (Note that we treat small-$N$ encounters up to $N=4$
via direct integration.)  We therefore truncate the mass function at $18.5\, M_\sun$ 
(the largest progenitor mass not resulting in a BH) for the runs presented here.
The resulting discrepancy in $r_c/r_h$ is important only at early times and
as Fig.~\ref{fig:hurleycomp} shows is minimal.

\begin{figure}
  \begin{center}
    \includegraphics[width=0.9\columnwidth]{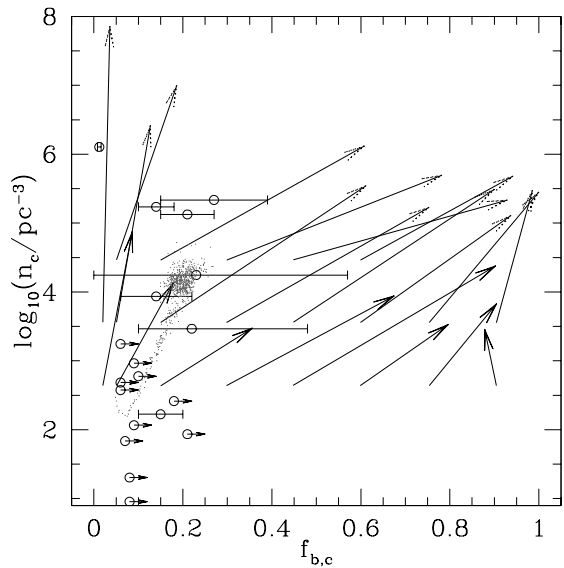}
    \caption{Evolution of our Monte Carlo cluster models in 
      core number density---binary fraction space.  Conventions are as
      in Fig.~\ref{fig:ivanovacomp}.  Solid arrowheads represent values
      measured at 14 Gyr, while dotted arrowheads are values measured
      before tidal disruption (since these models didn't last for 14 Gyr) 
      at times between $\sim 8$ and $\sim 13 \, {\rm Gyr}$ for the medium initial
      density models, and between $\sim 2$ and $\sim 12 \, {\rm Gyr}$ for the
      high initial density models.  For reference, the detailed evolution of
      the low-density $f_b=0.05$ model is shown in small gray dots.  The low initial
      density models have initial half-mass relaxation times of $t_{\rm rh}=0.8\, {\rm Gyr}$,
      the medium density models have $t_{\rm rh}=0.3\, {\rm Gyr}$, and the high density
      models have $t_{\rm rh}=0.09\, {\rm Gyr}$.\label{fig:grid}}
  \end{center}
\end{figure}

We have performed several simulations of evolving clusters for a grid
in initial binary fraction and initial cluster virial radius (or equivalently, central
density).  All our simulations start with $N=10^5$ objects initially (an object being
either a binary or a single star), and like the simulations of 
\citet{2007ApJ...665..707H}, assume a Plummer density profile with 
no primordial mass segregation, a ``standard'' Galactic tide (cluster
at $8.5\,{\rm kpc}$ from Galactic center, $10^{11}\,M_\sun$ Galactic
mass enclosed), a \citet{1993MNRAS.262..545K} IMF, and only hard binaries.
Our IMF extends from $0.15$ to $18.5\,M_\sun$, binary secondary masses
are drawn from a distribution flat in the mass ratio, the semimajor
axis $a$ is drawn from a distribution flat in $\log a$ from a minimum
of $a_{\rm min}=5(R_1+R_2)$, where $R_i$ are the individual stellar radii, 
to a maximum corresponding to an orbital velocity of the lighter member equal
to the local velocity dispersion, and the eccentricity
is drawn from a thermal distribution truncated at the value corresponding
to contact at $a_{\rm min}$.  Note that our large $a$ cutoff for wide binaries is equivalent
to the hard--soft boundary for equal-mass stars \citep{2006ApJ...640.1086F}.

Fig.~\ref{fig:grid} shows the evolution of our models in core number 
density---binary fraction space.  It is clear that for all the but the highest
initial binary fraction cases, the core binary fraction increases with time.  
The observational data points seem to be consistent only with cluster models
that started with relatively low central densities ($\sim 10^3 \, {\rm pc}^{-3}$)
and small hard binary fractions ($\sim 5 \%$).  As we discuss in the next section,
the core binary fraction is typically estimated observationally by measuring the fraction
of main sequence stars belonging to the binary main sequence, and convolving it with 
an assumed binary mass ratio distribution.  It is not a priori evident that this MS
binary fraction reflects the underlying true binary fraction.

\begin{figure}
  \begin{center}
    \includegraphics[width=0.9\columnwidth]{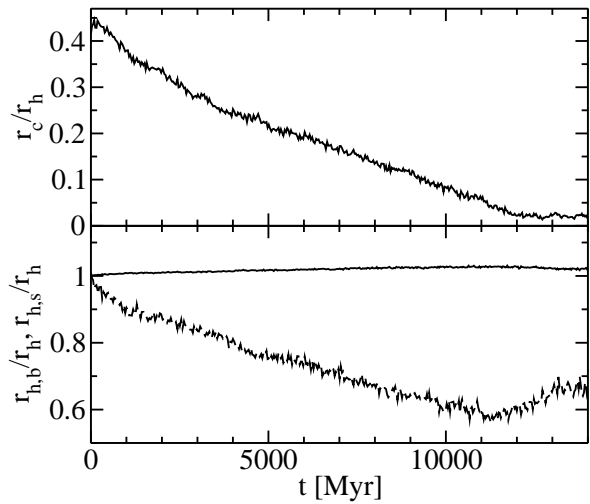}
    \caption{Evolution of our Monte Carlo cluster model
      starting with $f_b=0.05$ and $n_c \approx 10^{2.5} \, {\rm pc}^{-3}$.
      The top panel shows the evolution of $r_c/r_h$ with time.  The cluster
      enters a binary burning phase at $\sim 12 \, {\rm Gyr}$.
      The bottom panel shows the evolution of the half-mass radius of
      single stars, $r_{\rm h,s}$ (solid line), and the half-mass
      radius of binaries, $r_{\rm h,b}$ (dashed line), relative to the
      overall cluster half-mass radius, $r_h$.  The differential mass
      segregation between the single and binary populations is evident, with
      the single stars expanding slightly relative to the bulk of the cluster,
      and the binaries contracting significantly. The quantity
      $r_{\rm h,b}/r_h$ decreases steadily until $\sim 11 \, {\rm Gyr}$ due
      to mass segregation, at which point it begins to increase due to destruction
      of binaries preferentially in the cluster core.\label{fig:rh}}
  \end{center}
\end{figure}

\begin{figure}
  \begin{center}
    \includegraphics[width=0.9\columnwidth]{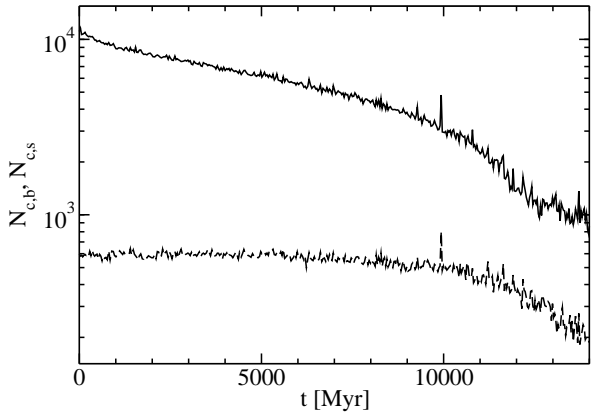}
    \caption{Evolution of the number of single stars in the core, $N_{\rm c,s}$ (solid line), and number of
      binaries in the core, $N_{\rm c,b}$ (dashed line), in our Monte Carlo cluster evolution model
      starting with $f_b=0.05$ and $n_c \approx 10^{2.5} \, {\rm pc}^{-3}$.
      The quantity $N_{\rm c,s}$ declines steadily due to standard gravothermal evolution,
      in which the cluster core becomes denser and smaller in number with time.
      The quantity $N_{\rm c,b}$ is roughly steady until $\sim 11 \, {\rm Gyr}$
      due to mass segregation of binaries into the cluster core.\label{fig:nc}}
  \end{center}
\end{figure}

Why does the core binary fraction generally tend to increase with time?  As mentioned above,
there are many strongly coupled processes that affect the core binary fraction.
However, the general trend can be understood approximately as an interaction between mass segregation
of binaries into the core, and the destruction of binaries preferentially in the core.

Fig.~\ref{fig:rh} shows the evolution of our Monte Carlo cluster evolution model
starting with $f_b=0.05$ and $n_c \approx 10^{2.5} \, {\rm pc}^{-3}$.
As the evolution of $r_c/r_h$ in the top panel shows, the cluster core contracts
steadily until it enters a phase of binary burning at the relatively late time of 
$\sim 12 \, {\rm Gyr}$.
The bottom panel shows the evolution of the half-mass radius of
single stars, $r_{\rm h,s}$ (solid line), and the half-mass
radius of binaries, $r_{\rm h,b}$ (dashed line), relative to the
overall cluster half-mass radius, $r_h$.  The differential mass
segregation between the single and binary populations is evident, with
the single stars expanding slightly relative to the bulk of the cluster,
and the binaries contracting significantly. The quantity
$r_{\rm h,b}/r_h$ decreases steadily until $\sim 11 \, {\rm Gyr}$ due to mass 
segregation.  It then begins to increase due to destruction of binaries 
preferentially in the cluster core by strong dynamical interactions and perturbed stellar evolution
\citep[see e.g.,][for a discussion of perturbed binary evolution]{2005MNRAS.358..572I}.

Fig.~\ref{fig:nc} shows the evolution of the number of single stars in the core, $N_{\rm c,s}$ 
(solid line), and number of binaries in the core, $N_{\rm c,b}$ (dashed line) for the same model.
The quantity $N_{\rm c,s}$ declines steadily due to standard gravothermal evolution,
in which the cluster core becomes denser and smaller in number with time 
\citep[e.g.,][]{2008gady.book.....B}.
The quantity $N_{\rm c,b}$, on the other hand, is roughly steady until $\sim 11 \, {\rm Gyr}$
due to mass segregation of binaries into the cluster core.

As suggested by Fig.~\ref{fig:nc}, the core mass decreases with time, as
expected from standard gravothermal evolution.  This is in contrast with the 
simplified Monte Carlo method of \citet{2005MNRAS.358..572I}, in which the core mass steadily
increases in time, due primarily to mass segregation of binaries into a core of fixed density.

We note also that mass segregation of a binary into the core implies, by energy 
conservation, a preferential expansion of lighter single stars in the vicinity of the 
binary.  (Energy conservation is roughly applicable because the mass segregation
timescale is shorter than the local relaxation timescale, by a factor of $M/m$, where
$M$ is the mass of the segregating object and $m$ is the mass of a background star.)
This effect is not included in the code of \citet{2005MNRAS.358..572I},
and is likely an important factor in the discrepancy between their results and ours.

Another important factor, as suggested by Figs.~\ref{fig:hurleycomp} and \ref{fig:rh}, is 
that the long lived, high density binary burning phase assumed by \citet{2005MNRAS.358..572I}
may not be generic for globular clusters.  Instead, the ``core contraction'' phase 
may last a Hubble time, and the cluster cores we observe now may have been
much less dense in the past \citep{2008ApJ...673L..25F}.  Although the central density 
in our models increases steadily with time, the local density at the half-mass radius {\em
decreases} with time, resulting in final half-mass relaxation times that are a factor
of $\sim 3$ {\em longer} than their initial values.  In the cases where our models do enter
the binary burning phase before a Hubble time, we find that the core binary fraction in
this phase steadily decreases with time.  This behavior is consistent with the results of
\citet{2005MNRAS.358..572I}.

While the core binary fraction in the majority of our models increases with time, the 
overall cluster binary fraction remains roughly constant with time.  This is in good agreement
with the \citet{2008AJ....135.2129H} $N$-body models inspired by NGC 6397, and supports their use of
the currently observed binary fraction near the half-mass radius as a measure of the
{\em primordial} binary fraction (although the validity of comparison with NGC 6397 is not obvious,
since the $N$-body models end with a factor of 5 to 10 fewer stars than NGC 6397 currently contains).  
For the low-density $f_b=0.05$ model just 
described, 39\% of the initial binary population remains at 14 Gyr, 43\% escape 
the cluster due to the tidal field (compared with the 60\% of single stars that 
escape in the same fashion), 9\% are destroyed via strong dynamical interactions
of binaries, and 8\% are destroyed via binary stellar evolutionary processes (possibly
perturbed by dynamics).  In other words, in this case the overall binary fraction remains
roughly constant with time due to a balance between preferential tidal stripping of
single stars in the outskirts and preferential destruction of binaries in the cluster core.

\section{Hiding Binaries}\label{sec:hidingbinaries}

When using the offset main sequence method, what observers measure is in
fact the number of MS--MS binaries with mass ratios $q \gtrsim 0.5$ relative
to the total number of objects appearing in the main sequence (which may 
include {\em apparent} single MS stars, comprised of a MS star plus
dim compact object companion).  This fraction is then corrected
to account for the low mass ratio MS--MS binaries that blend into the single MS, 
by adopting an assumed mass ratio distribution.  This final corrected
figure is what is usually quoted as the ``observed binary fraction.''
However, there is no a priori reason to believe this quantity reflects the 
underlying binary fraction among stars of {\em all} types.  \citet{2008AJ....135.2129H}
showed that, for the low binary fraction cluster models they considered ($f_b \lesssim 10\%$),
the observed binary fraction is a good measure of the true binary fraction in the outer
regions of a cluster, but can be a serious overestimate in the core.  

\begin{figure}
  \begin{center}
    \includegraphics[width=0.9\columnwidth]{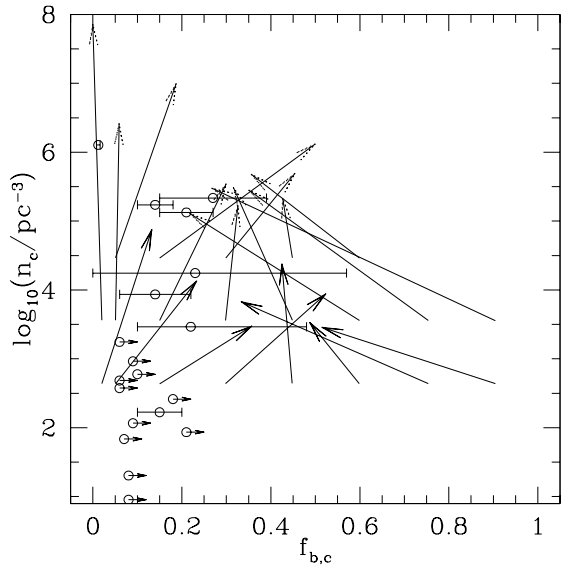}
    \caption{Same as Figure \ref{fig:grid}, but for main sequence binaries.\label{fig:gridobs}}
  \end{center}
\end{figure}

Since the general nature
of the relationship between the observed and true binary fraction is not obvious, we have plotted
in Fig.~\ref{fig:gridobs} the evolution of our models in core number density--{\em observed}
core binary fraction space.  The observed binary fraction is calculated as 
$N_{\rm MS-MS}/(N_{\rm MS-MS}+N_{\rm MS}+N_{\rm MS-CO})$, where $N_{\rm MS-MS}$
is the number of MS--MS binaries of any mass ratio in the core, $N_{\rm MS}$ is the number
of single MS stars in the core, and $N_{\rm MS-CO}$ is the number of MS--compact object
binaries in the core that appear near the MS.  We count a MS--compact object binary as near the MS
if the total luminosity of the binary is less than 10\% more than that of the MS star (this corresponds
to a magnitude increase of 0.1), and if the luminosity-weighted temperature of the binary is less
than 10\% different from that of the MS star.  Like the true binary fraction plotted in Fig.~\ref{fig:grid}, small
initial binary fraction models evolve toward larger binary fractions.  However, large initial
binary fraction models evolve toward drastically smaller observed binary fractions.  As 
a relatively extreme example, the model with an initial binary fraction of 75\% and initial 
central density of $\sim 10^{2.5}\, {\rm pc}^{-3}$ has an observed core binary fraction of 
just 33\% at the end of the simulation.  The true core binary fraction at
the end of the simulation is 91\%.  Of the core binaries, 23\% are MS--MS binaries,
32\% are compact object--compact object binaries, and 44\% are MS--compact object
binaries (see Table \ref{tab:corebins} for more details).  
As expected, the discrepancy between the observed and the true binary fraction
is due to compact object--compact object binaries not being counted in the observed
tally, and MS-compact objects masquerading as single stars on the main sequence.

\begin{deluxetable}{ccc}
  \tablecaption{Population breakdown of core binaries at 14 Gyr for the $\sim 10^{2.5}\, {\rm pc}^{-3}$
    initial core density, 75\% initial binary fraction model.\label{tab:corebins}}
  \tablehead{
    \colhead{type} & \colhead{number} & \colhead{fraction}
  }
  \startdata
  MS--WD & 139 & 44\%\\
  WD--WD & 101 & 32\%\\
  MS--MS & 74 & 23\%\\
  NS--WD & 1 & 0.3\%\\
  HG--WD & 1 & 0.3\%
  \enddata
  \tablecomments{The third column is the fraction of the total number of core binaries represented
  by that binary type.  ``MS'' denotes main sequence, ``WD'' denotes white dwarf, ``NS'' denotes
  neutron star, and ``HG'' denotes Hertzsprung gap star.}
\end{deluxetable}

\section{The Importance of Soft Binaries}\label{sec:softbinaries}

An initial population of binaries that contains a substantial soft component can be
a significant cluster energy {\em sink}, since the soft binaries are destroyed in 
dynamical scattering interactions.  The result is that the core of a cluster born 
with many soft binaries will quickly contract as those binaries are ionized.
Could soft binaries increase the concentration of a cluster so much that it would
become core collapsed?  

The total energy in soft binaries, for a distribution 
flat in the log of the semimajor axis, is simply
\begin{equation}
  E_{\rm b,s} = \frac{N_b (E_{\rm b,hs} - E_{\rm b,amax})}{\ln(a_{\rm max}/a_{\rm min})}
                 \approx \frac{N_b E_{\rm b,hs}}{\ln(a_{\rm max}/a_{\rm min})} \, ,
\end{equation}
where $a_{\rm min}$ and $a_{\rm max}$ are the limits on the semimajor axis distribution,
$E_{\rm b,hs}$ is the energy of a binary at the hard--soft boundary, 
$E_{\rm b,amax}$ is the energy of the least-bound binary, and $N_b$ is the total
number of binaries.  Assuming for simplicity a cluster of equal-mass objects (binaries
in this case) of mass $m_{\rm ave}$ with mean 1D velocity dispersion $\sigma$, this becomes
\begin{equation}
  E_{\rm b,s} \approx \frac{\frac{3}{2} N_b m_{\rm ave} \sigma^2}{\ln(a_{\rm max}/a_{\rm min})} \, .
\end{equation}
From the virial theorem, the total mechanical energy of a cluster is simply 
$E_{\rm clus} = -\frac{3}{2}Nm_{\rm ave}\sigma^2$, where $N$ is the number of cluster objects, so
\begin{equation}
  \frac{E_{\rm b,s}}{|E_{\rm clus}|} \approx \frac{N_b}{N \ln(a_{\rm max}/a_{\rm min})} \, .
\end{equation}
For an admittedly optimistic binary fraction of 1 ($N_b=N$), and realistic binary
semimajor axis limits of $a_{\rm min} = 5 \times 10^{-2} \, {\rm AU}$ (corresponding
to a contact binary during the pre-main sequence phase) and 
$a_{\rm max}=10^{3} \, {\rm AU}$ (corresponding to a $10^7$ day orbital period),
the energy in soft binaries is $\approx 10\%$ of the total cluster mechanical
energy!

The question, of course, is if this amount of energy is sufficient to make a cluster
concentrated enough to appear to be core collapsed.  For our working definition
of core collapse we assume that a cluster core can be resolved with HST if its
radius is at least 1 arcsecond in size.  
At a typical cluster distance of 10 kpc, this corresponds to $\sim 0.05 \, {\rm pc}$.
Starting with a King model of a given mass, binary fraction, central concentration $W_0$,
and half-mass radius $r_h$, we calculate the total mechanical energy of the cluster within the half-mass
radius, $E_h$ \citep{2008gady.book.....B}.  We then calculate the energy of the soft binaries, $E_{\rm b,s}$.
This energy will be absorbed from the cluster when those binaries are destroyed in dynamical interactions 
in and around the cluster core.  Keeping $r_h$ fixed (since the timescale for destruction of soft binaries
is shorter than the half-mass relaxation time), we then calculate a new King model
with half-mass energy $E_h^\prime = E_h-E_{\rm b,s}$ (note that $E_{\rm b,s} > 0$ by construction).
For the new King model we calculate the new central velocity dispersion and hence the new hard--soft
boundary (which has moved to smaller binary semimajor axis), calculate the energy available in the 
newly soft binaries, and iterate until the solution converges.
For a $5 \times 10^5 \, M_\sun$ cluster with half-mass radius $r_h=5 \, {\rm pc}$ 
and initial core radius $r_c=1.9 \, {\rm pc}$ ($W_0=6$, concentration $c=\log_{10}(r_t/r_c)=1.25$), 
an initial binary fraction of 100\% with semimajor axis distributed flat in $\log a$ from $5 \times 10^{-2} \, {\rm AU}$
to $10^3 \, {\rm AU}$ is sufficient to drive the cluster to a $W_0=10$, $c=2.3$ King model
with core radius $r_c=0.16 \, {\rm pc}$.  (A $W_0=10$ King model has maximal binding energy
within $r_h$ for fixed $r_h$ and mass.)  This is quite close to core collapsed, and may even
be classified as such if viewed with a ground-based telescope.  In fact, \citet{1985PASJ...37..715W}
showed that clusters enter the self-similar stage of evolution (the ``onset'' of core collapse) 
when $W_0 > 7.4$, so such a model would reach core collapse quickly.  We have repeated this calculation
with a binary distribution that is log-normal in orbital period, as in \citet{1991A&A...248..485D} or 
\citet{1992ApJ...396..178F}, with $\langle \log_{10} P_d \rangle = 4.8$ and
$\sigma_{\log_{10} P_d}=2.3$, where $P_d$ is the period in days, and with the same limits on 
semi-major axis as above.  The results are unchanged with this binary distribution, largely because 
its peak lies at wider orbits than the hard--soft boundary for globular clusters.

\begin{figure}
  \begin{center}
    \includegraphics[width=0.9\columnwidth]{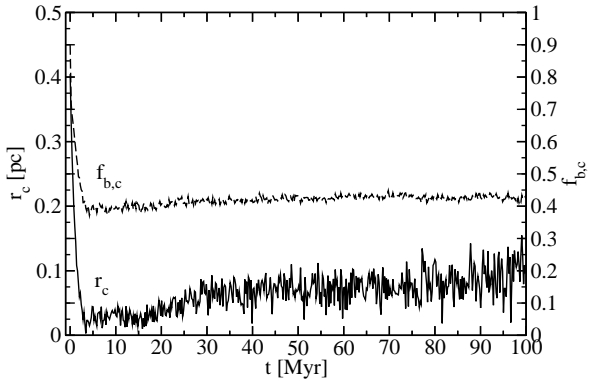}
    \caption{Evolution of the core radius and core binary fraction for our ``high density'' initial
      model with a 90\% initial binary fraction, including soft binaries.  The core contracts rapidly
      at the start of the evolution due to the destruction of soft binaries, and quickly enters
      the binary burning phase.\label{fig:soft}}
  \end{center}
\end{figure}

To test this scenario numerically, we have run models with a binary distribution extending well beyond
the hard--soft boundary, to $P = 10^7 \, {\rm d}$.  Our ``high density'' model (cluster 
mass $9 \times 10^4 \, M_\sun$, standard wide mass spectrum, $r_h=1.0 \, {\rm pc}$ 
initially) with $f_b=0.9$ (including soft binaries), evolves from $f_{\rm b,c}=0.9$ and 
$r_c=0.6 \, {\rm pc}$ to $f_{\rm b,c}=0.4$ and 
$r_c=0.05 \, {\rm pc}$ in just 3 Myr (see Fig.~\ref{fig:soft}).  This is in striking agreement with the energy argument
above, which predicts rapid evolution to $r_c=0.1 \, {\rm pc}$ for this model.  Note that
the energy argument assumes {\em all} soft binaries will be destroyed on a short timescale.
To achieve this in practice requires efficient mass segregation of binaries into the core, which
has been aided in this case by a wide mass spectrum, at the expense of inaccuracy in calculating
$E_{\rm b,s}$.  After the rapid initial contraction of the core, the cluster quickly
(after a few Myr) enters into a long-lived binary burning phase.  

From the preceding discussion, it is evident that the dynamical importance of soft binaries
should not be ignored.  If a cluster is born with significant numbers of soft binaries, its
evolution may be vastly different from a similar cluster containing only hard binaries.
First, the rate of binary destruction is greatly enhanced in clusters containing soft binaries, yielding a 
starkly decreasing binary fraction with time.  Second, the binary burning phase is reached 
quickly (within a few Myr) due to soft binary destruction.  When only hard
binaries are present, the binary burning phase may not be reached within a Hubble time, as 
shown for example by Fig.~\ref{fig:hurleycomp}.  The implications for our understanding of
the current dynamical states of Galactic globular clusters are profound, as certain
properties of clusters can be explained by the majority of clusters currently being in the initial ``core
contraction'' phase, and not yet in binary burning \citep{2008ApJ...673L..25F}.  We have 
provided here just a cursory analysis  of the effects of soft binaries.  A more detailed study 
should certainly be undertaken in the future.

\section{Discussion}\label{sec:disc}

Independent of the details, it seems clear that the hard binary fraction in the core
of a dense stellar system will generally increase with time (with the exception of 
an initial hard binary fraction $\gtrsim 90\%$).  Yet there is no compelling evidence
that clusters should be born with binary fractions smaller than the typical field value
of $\sim$ 50\%, and observations yield core binary fractions of just $\sim$ 10\% 
in Galactic globular clusters.  If the observations are to be taken at face value, how then can they
be consistent with large initial binary fractions?  One possibility, as pointed out
by \citet{2008AJ....135.2155D}, is that the binary fraction is a strong function of 
primary mass \citep{2006ApJ...640L..63L}, with the {\em single} star fraction increasing to
$\sim 75\%$ for M dwarfs and lighter stars.  A \citet{1993MNRAS.262..545K} IMF
with a 25\% binary fraction from 0.1 to 0.5 $M_\sun$ and a 50\% binary fraction from
0.5 to 100 $M_\sun$ yields an overall binary fraction of just 32\%.  

Another possibility is that most binaries born in clusters are soft relative
the cluster velocity dispersion, in which case they will be destroyed very
quickly by dynamics.  If the binary period distribution is uniform 
in $\log P$ from 0.1 to $10^7$ d as in \citet{2005MNRAS.358..572I}, the 32\% overall binary fraction 
just suggested corresponds 
to a hard binary fraction of merely $\sim 10\%$ for a cluster with central density
$10^6 \, {\rm pc}^{-3}$.  As demonstrated above, the early, rapid destruction
of soft binaries may lead to a binary-burning phase within a short time
($\lesssim 5 \, {\rm Myr}$, depending on initial conditions).  

Aside from the initial binary properties, could it be that observations are under-counting
the binary fraction significantly?  When we measure the binary fraction using an offset 
main sequence method similar to what observers use, we find that clusters with large initial binary
fractions ($f_b \gtrsim 0.5$) evolve toward smaller {\em observed} core 
binary fractions ($f_b \lesssim 0.5$).  The discrepancy between the observed and true core binary
fractions is caused by compact object-compact object binaries not being counted in the sample,
and MS-compact object binaries masquerading as single stars.  

Could a binary be sufficiently wide
to be resolved as two single stars and hence missed as a binary?
For the wide-field camera on HST, one requires two turnoff mass stars in a binary to be separated by
roughly 4 pixels for the binary to be resolved.  For a cluster at a distance of
10 kpc, this corresponds to a binary separation of $\sim 4\times 10^3 \, {\rm AU}$.
For a cluster with velocity dispersion $10 \, {\rm km} \, {\rm s}^{-1}$, this corresponds
to a binary hardness of $Gm/a v_\sigma^2 \approx 2 \times 10^{-3}$, which is too soft 
to survive dynamically for even a short time.

\section{Summary}\label{sec:summary}

We have described in detail our inclusion of the BSE single and binary stellar evolution routines in
our Monte Carlo globular cluster evolution code \citep{2000MNRAS.315..543H,2002MNRAS.329..897H}.  We 
have compared with the results of direct $N$-body simulations and found good agreement, suggesting that 
our implementation of BSE in our code is consistent with that in $N$-body.

We have used our newly upgrade Monte Carlo code to study the evolution of the core hard binary
fraction in star clusters, and in particular attempt to settle the apparent disagreement
between direct $N$-body and simplified Monte Carlo techniques on its evolution.
We find that the core binary fraction generally increases with time, even
for low initial core density models ($n_c \approx 10^{2.5} \, {\rm pc}^{-3}$), with
only very small initial binary fraction models ($f_b \lesssim 0.05$) producing the
core binary fractions of $\sim 10\%$ observed today.  The increase in the core binary
fraction with time can be understood as an imbalance between mass segregation of
binaries into the core (and single stars out of the core) and the destruction of
binaries in the core directly via strong dynamical encounters, and indirectly via dynamical
perturbation of binary stellar evolution processes.  The overall cluster binary fraction
remains roughly steady with time, due to the additional effect of preferential tidal
stripping of single stars from the cluster outskirts.

This evolution, however, refers to the {\em true} binary fraction.  When measuring 
the core binary fraction using
an offset main-sequence method analogous to what observers use, we find that the
{\em observed} core binary fraction can seriously underestimate the true core
binary fraction.  This results from compact object-compact object binaries not 
being counted and MS-compact object binaries masquerading as single stars
in the observed tally.
In the course of creating more detailed models of M 67, 47 Tuc, M 4, and NGC 6397
to be compared with observations, we are now developing a data
reduction pipeline that includes simulations of spectra for every star.  Among our
near future plans is the creation of a cluster sky map in different bands, to which we
can apply the MS binary detection method to determine more accurately how many binaries are 
missed by the method.

Most of our discussion concerned hard binaries.  
However, we also considered the effects of a substantial population of soft binaries.  We found
that the energy {\em sink} represented by soft binaries (for a typical binary period distribution) 
is sufficient to cause the core of a typical globular cluster to contract significantly.  The result
is not only a rapid, efficient destruction of a significant number of binaries at early times, but also a
much earlier onset of the binary burning phase, resulting in enhanced binary destruction in the core
with time.

\acknowledgements

The authors thank C. Heinke and J. Hurley for data and helpful discussions.
JMF acknowledges support from Chandra/Einstein Postdoctoral Fellowship Award PF7-80047.
FAR acknowledges support from NASA Grant NNG06GI62G at Northwestern University.
This research was completed at KITP while the authors participated in the 
spring 2009 program on ``Formation and Evolution of Globular Clusters,'' and was 
supported in part by the NSF under Grant PHY05-51164.

\bibliographystyle{apj}
\bibliography{apj-jour,main}

\clearpage

\end{document}